\documentclass[journal=jctcce,manuscript=article]{achemso}

\usepackage[version=3]{mhchem} 
\usepackage{mciteplus}

\usepackage{bm}
\usepackage{physics}
\usepackage{graphicx}
\usepackage{subcaption}
\usepackage{xcolor}

\newcommand{\RR}{\bm{R}}
\newcommand{\dg}[1]{{#1}^{\dagger}}
\newcommand{\ti}{\tilde{i}}
\newcommand{\ta}{\tilde{a}}
\newcommand{\tj}{\tilde{j}}
\newcommand{\tb}{\tilde{b}}

\author{Zuxin Jin}
\author{Joseph E. Subotnik}

\title{Nonadiabatic dynamics at metal surfaces: fewest switches surface hopping with electronic relaxation}

\begin{document}
	
	\begin{abstract}
		A new scheme is proposed for modeling molecular nonadiabatic dynamics near metal surfaces. 
		The charge-transfer character of such dynamics is exploited to construct an efficient reduced representation for the electronic structure. In this representation, the fewest switches surface hopping (FSSH) approach can be naturally modified to include electronic relaxation (ER).
		The resulting FSSH-ER method is valid across a wide range of coupling strength as supported by tests applied to the Anderson-Holstein model for electron transfer. Future work will combine this scheme with \textit{ab initio} electronic structure calculations.
	\end{abstract}
	
	\section{Introduction} {
		Molecular nonadiabatic dynamics near metal surfaces
		has attracted widespread interest across many areas, including gas-phase scattering\cite{Wodtke2004, Bartels2011, Kandratsenka2018}, molecular junctions\cite{Alemani2006,Danilov2006,Donarini2006,Henningsen2007}, and dissociative chemisorption\cite{Jiang2016, Maurer2017, Yin2018, Jiang2018}.
		Because low-lying electron-hole pairs (EHPs) can be excited so easily in a metal, such dynamics can easily go beyond the Born-Oppenheimer approximation, as indicated by various phenomena like chemicurrents\cite{Nienhaus1999, Gergen2001}, unusual vibrational relaxation\cite{Wodtke2000, Kruger2016, Wagner2017, Kumar2019} and inelastic scattering\cite{Bunermann2015, Dorenkamp2018, Steinsiek2018}.
		Therefore, to fully understand these processes, a robust approach to nonadiabatic dynamics would be extremely useful.
		Nevertheless, in spite of the enormous progress made to date\cite{Persson1980, Andersson1981, Persson1982, Tully2009I, Tully2009M, Shenvi2009S, Dou2016BCME, Miao2017, Rittmeyer2017, Meyer2018, Elste2008, Bode2011, Bode2012, Dzhioev2013, Galperin2007, Galperin2015, Chen2019, Chen2019a, Scholz2016, Juaristi2017, Loncaric2017, Bouakline2019, Scholz2019, Fischer2020}, modeling such dynamics remains a very difficult task and poses tough problems for both electronic structure calculations and dynamics simulations.
		
		In terms of the electronic structure, the heterogeneous nature of a molecule-metal interface raises a basic question: what is the appropriate representation for describing a typical molecule-metal system? 
		In particular, two sub-questions must be addressed:
		
		First, should we adopt a simple picture of independent (or mean-field) electrons or a more complex picture of interacting electrons? The former is far more efficient than the latter and is known to be adequate in many systems. However, it is dubious whether an independent electron picture is sufficient for the majority of reactions. After all, for molecules alone (i.e. without a metal), a high-level electronic structure method is usually necessary if we are to model bond-making and bond-breaking\cite{Noga1987, Bartlett2007, Sherrill2005}: why should the presence of a metal makes the problem that much simpler?
		
		Second, note that while the entire system possesses a large number of electronic degrees of freedom(DoFs), the number of distinct molecular electronic states is always much smaller. 
		Thus, if one is only interested in the molecular dynamics, one can ask: is it necessary to work with the entire system's adiabatic states (or one-electron eigenstates), or can we safely seek a reduced representation? A reduced picture would be far more computationally attractive/feasible. However, what would be a good enough reduced picture? Would a simple molecular diabatic representation suffice, or must we seek a more accurate alternative? 
		
		Next, let us turn to nuclear dynamics. In a fully quantum picture, when a nuclear wave packet passes through a crossing with non-vanishing coupling, the wave packet splits into individual wave packets each associated with individual electronic potential energy surfaces(PESs). This picture lies at the heart of nonadiabatic dynamics, and it remains valid at both high and low temperatures. However, because of the formidable cost of simulating quantum nuclei and the fact that \(kT \) in many scenarios is larger than the characteristic energy of low-frequency nuclear motion, semi-classical approaches have become popular today.  Quite often, a nuclear wave packet and its splitting near a crossing is modeled by an ensemble of classical trajectories and their branching. Nevertheless, in order to achieve a quantitative description, many approximations will be necessary, some of which are uncontrolled and will need to be analyzed against accurate benchmark studies. For the present paper, we will assume that a classical simulation of nuclei is sufficient, and we will focus on all of the other problems that arise as far as nonadiabatic effects.
		
		When simulating nuclear dynamics at a metal surface, it cannot be emphasized enough both (i) that capturing the dynamics accurately requires modeling many electronic states to represent the continuum and (ii) that modeling so many electronic states can be a computational quagmire.	
		This situation can and should be contrasted with the case of nonadiabatic dynamics involving only a handful of discrete electronic states, e.g. a gas-phase photo-excited molecule, where there are many tools for solving for electronic structure and studying nonadiabatic dynamics.
		For instance, for a molecule in the gas phase, accurate excited potential energy surfaces can often be achieved by time-dependent density functional theory (TDDFT) or multi-configuration self-consistent field (MCSCF) methods; if necessary, adiabatic-to-diabatic transformations can be performed to generate a diabatic picture; and finally, a variety of (nearly) exact dynamical schemes have been proposed, including the Miller-Meyer-Stock-Thoss approach\cite{Meyer1979, Stock1997}, multi-configuration time-dependent Hartree\cite{Meyer1990, Manthe1992, Beck2000, Wang2003, Wang2009}, multiple spawning\cite{BenNun1998}, hierarchical quantum master equation\cite{Tanimura1989,Tanimura1990, Tanimura2006, Ishizaki2005, Ishizaki2009,Ishizaki2009a}, quantum Monte-Carlo\cite{Rabani2008, Chen2016, Chen2017, Chen2017a}, linearized density matrix dynamics\cite{Huo2010, Huo2011}, etc. {While many of these methods do in principle allow an arbitrary number of electronic DoFs and some of them have been applied to systems with a fermionic bath\cite{Thoss2007, Rabani2008, Chen2016, Chen2017, Chen2017a, Schinabeck2016, Xu2019}, it remains a difficult task to treat a realistic \textit{ab initio} Hamiltonian with a continuum of electronic states in the presence of a large number of anharmonic vibrational modes.}
		For the present problem, the number of energetically relevant excited states is prohibitively large due to the possibility of exciting low-lying EHPs. 
		
		Beyond the requirement of handling a large number of electronic states, another difficulty
		when modeling nuclear-electronic dynamics at a metal surface is the need for accurate dynamics
		across a wide range of parameters. 
		For a molecule near a metal surface, there are at least three characteristic energy scales: the temperature \(kT\), the hybridization function \(\Gamma\), and the characteristic energy scale of nuclear motion \( \hbar\omega \). 
		Assuming \(kT \gg \hbar \omega\) so that nuclear motion can be viewed classical, there are well-established methods that apply to different regions of \(\Gamma\) as illustrated by Fig. \ref{fig:methods}\cite{Dou2016BCME}. 
		Specifically, for \(\Gamma < kT \), one arrives at a classical master equation, in which the effect of the metal can be captured by stochastic hops between molecular diabatic surfaces\cite{Weick2008, Dou2015II, Dou2015III}; for \(\Gamma > \hbar \omega\), it has been shown that nonadiabatic effects on nuclear dynamics can be well incorporated into the electronic friction tensor\cite{Hynes1993, Tully1995, Juaristi2008, Dou2015F, Rittmeyer2015, Dou2016, Tully2016prl, Tully2016, Novko2016, Dou2017prl, Dou2017F, Rittmeyer2017, Meyer2018, Dou2018perspective, Dou2018F}, leading to a Langevin dynamics on a potential of mean force\cite{Dou2017prl}.
		
		
		However, nuclear dynamics in the above two regions are not directly compatible with each other (though, see our description of the BCME in Sec. \ref{subsec:exist} below). Moreover, for realistic systems, both scenarios are possible, and, more often than not, a system may sample both regions in a single experiment. 
		Thus, 
		a method that works in both limits is needed.
		
		\begin{figure}
			\includegraphics[width=3.3in]{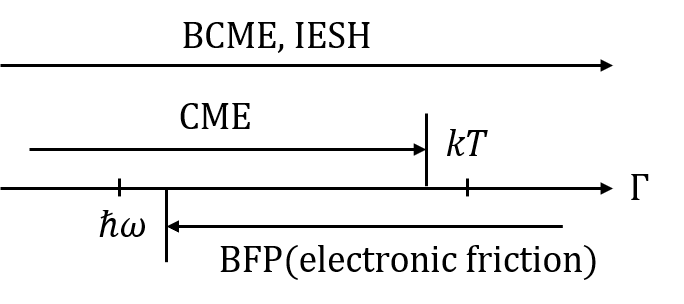}
			\caption{The coverage of the classical master equation (CME) and the broadened Fokker-Planck (BFP) method with respect to the hybridization function, assuming \(kT \gg \hbar \omega\). This figure is adapted from Fig. 1 of Ref. \citenum{Dou2016BCME}}
			\label{fig:methods}
		\end{figure}
	
		\subsection{Existing semi-classical approaches to nonadiabatic dynamics at metal surfaces}\label{subsec:exist} {
  			To our knowledge, there are today very few  algorithms for simulating nonadiabatic dynamics near a metal surface semiclassically that should be valid across a wide range of parameter regimes. Below we will give a brief review of two of them.
  			
  			The first method is the independent electron surface hopping (IESH)\cite{Tully2009I, Tully2009M, Roy2009OpenShell}, which is a variant of the famous fewest-switches surface hopping (FSSH)\cite{Tully1990}. 
			According to IESH, one allows individual electrons to hop between orbitals, eventually gathering statistics. As such, a single Slater determinant can capture a vast number of excited states. 
			In practice, IESH has been applied to the NO-Au scattering experiment\cite{Shenvi2009S} and often gives qualitatively good results. 
			Nevertheless, by definition, the independent electron picture cannot be extended to systems whose electronic structures are correlated beyond a mean-field approximation.
			
  			A second approach to this same problem is
			the broadened classical master equation (BCME)\cite{Dou2016BCME, Dou2017BCME}, which was recently developed by our research group and compared with IESH\cite{Miao2019}. BCME extrapolates the CME to the strong molecular-metal coupling regime (\(\Gamma>kT\)). The BCME yields accurate and efficient results for the Anderson-Holstein model\cite{Ouyang2016, Miao2019} and has been recently applied to electrochemical model problems\cite{Coffman2020}. However, at bottom, the BCME method is formulated in a (modified) molecular diabatic picture, which has both upsides and downsides. The upside is that, by construction, the BCME is very inexpensive because it does not need to treat a large number of electronic states explicitly (as opposed to IESH where all one-electron eigenstates are explicitly involved). The downside is that the method is not easily applied to realistic systems where one would like to perform \textit{ab initio} electronic structure calculations rather than estimate broadened diabats and a hybridization function that is forced to obey the wide-band approximation. 
		} 

		\subsection{Fewest Switches Surface Hopping with Electronic Relaxation (FSSH-ER)} {
			In this article, we will present another (third) method for 
			running nonadiabatic molecular dynamics at metal surfaces that will hopefully go beyond the methods described above and 
			be both computationally efficient as well as compatible with \textit{ab initio }electronic structure calculations; future publications will hopefully
			confirm these two assertions. Working in the context of the (non-interacting) Anderson-Holstein model, our specific approach will be as follows: First, we will start with a set of one-electron eigenstates (similar to IESH) 
			but then (unlike IESH) we will invoke a Schmidt decomposition to find pairs of Schmidt orbitals (one localized on the molecule, one localized on the metal). This pair of orbitals is analogous to the two molecular diabatic states that one would predict with a theory like the BCME.  Second, we will use these Schmidt orbitals to 
			construct an appropriate subspace of many-electron Slater determinants from which we can build a configuration interaction 
			Hamiltonian. Third and finally, we will apply a modified version of the FSSH to our system. Below, we will show that this nonadiabatic dynamics protocol is able to recapitulate Marcus's electrochemical theory (in the nonadiabatic limit) as well as transition state theory (in the adiabatic limit), which gives us hope that this new framework may be quite powerful going forward. The electronic structure and dynamics algorithms above should have natural extensions to \textit{ab initio} calculations beyond the Anderson-Holstein model.
			
			Regarding the outline of the article, in Sec. \ref{sec:method}, we present the theory described above, including the necessary choice of 
			electronic states and the proposed protocol for running nuclear-electronic dynamics. In Sec. \ref{sec:result}, we present results for the Anderson-Holstein model which describes electron transfer at a metal surface in an idealized fashion and is the basis of Marcus theory at a metal surface.  In Sec. \ref{sec:discussion}, we discuss our results, emphasizing a few nuances of the present algorithm (that may have gone unappreciated) and highlighting future numerical tests of the current protocol. We conclude in Sec. \ref{sec:conclusion}
		} 
		
	} 

	\section{Method}\label{sec:method} {
		In this work, we assume that the electronic Hamiltonian for the system of a molecule near a metal surface can be represented by the Anderson-Holstein model:
		\begin{align}\label{eq:H:QIM}
			H_e(\RR) = U_0(\RR) + \epsilon_d(\RR) \dg{d}d + \sum_{k} \epsilon_k \dg{c}_k c_k + \sum_k \left(V_k(\RR) \dg{d}c_k + V_k^*(\RR) \dg{c}_k d\right)
		\end{align}
		here, \(\RR \) represents the nuclear coordinate; \(d\) is the fermionic operator for the molecular orbital whose on-site energy is \(\epsilon_d \); \(U_0(\RR)\) and \(U_1(\RR) \equiv  U_0(\RR) + \epsilon_d(\RR) \) are the two molecular diabatic PESs; \( c_k \) is the operator for the metal state \(k\) whose energy is \(\epsilon_k \);  \(V_k\) is the hopping amplitude between the molecular orbital and the metal state \(k\). 
		
		The hybridization function \(\Gamma \) is defined to be
		\begin{align}\label{eq:hybrid}
			\Gamma(\epsilon) = 2\pi\sum_k \abs{V_k}^2 \delta(\epsilon-\epsilon_{k})
		\end{align}
		In the wide-band limit, \(\Gamma \) is independent of energy, and it represents the rate of electronic relaxation of the impurity orbital. Eq. \ref{eq:H:QIM} can represent a molecule-metal system only when \(\{\epsilon_{k} \}  \)  forms a quasi-continuum such that the energy spacing of \(\epsilon_{k} \) is smaller than any relevant characteristic energy scale (which includes \(\Gamma\), of course).
		
		This non-interacting Hamiltonian can be directly diagonalized:
		\begin{align}\label{}
			H_e(\RR) = U_0(\RR) + \sum_{p} \lambda_p(\RR)  \dg{b}_{p} {b}_p
		\end{align}
		and its ground state is a Slater determinant
		\begin{align}\label{eq:GS:nonint}
			\ket{\Psi_0} = \ket{ij \ldots}
		\end{align}
		where \(ij\ldots\) are occupied orbitals. 
		
		For a non-interacting Hamiltonian like Eq. \ref{eq:H:QIM}, the above ground state is exact. For realistic systems where there are electron-electron interactions, a ground state of the form of a Slater determinant can usually be constructed by assuming a mean-field ansatz -- which may or may not be valid. We will assume that Eq. \ref{eq:GS:nonint} is valid throughout this work, but see Sec. \ref{subsec:multi_ee} and Ref. \citenum{Chen2020} for our initial steps towards treating electron-electron interactions.
		
		In order to model nonadiabatic dynamics, electronic excited states must be considered. 
		While such electronically excited states can be delineated by counting all of the possible configuration states (\(\ket{\Psi_i^a}, \ket{\Psi_{ij}^{ab}}, \ldots \)), the total number of such states is enormous and forbids a direct application of conventional mixed quantum-classical methods like the FSSH.
		To address this problem, IESH\cite{Tully2009I} was introduced as a variant of the FSSH. By allowing electrons to hop individually between orbitals (instead of working with a many-electron basis), IESH effectively sample a vast number of possible configuration states. Nevertheless, this assumption also makes IESH difficult to extend to systems where electron-electron interactions and correlations become significant.
		Below, we will propose another variant of the FSSH. In particular, we will use a reduced many-electron description that focuses on charge-transfer nonadiabatic effects.
		
		\subsection{Construction of a Reduced Representation}\label{sec:redrep} {
			\subsubsection{Orbital Rotation} {
				An important fact about Slater determinants is that they are invariant under a unitary transformation of their orbitals. Given the definition of \(\ket{\Psi_0} \) in Eq. \ref{eq:GS:nonint}, let \(U \) be a unitary matrix, and \(\ket{\tilde{j}} = \sum_{i}^{occ} \ket{i} U_{ij} \) be a set of new orthonormal orbitals. Then \(\ket{\tilde{\Psi}_0} \equiv \ket{\tilde{i}\tilde{j}\ldots } = \det(U)\ket{\Psi_0} \) differs from the canonical ground state (Eq. \ref{eq:GS:nonint}) by merely a phase. In other words, \(\ket{\tilde{\Psi}_0} \) and \(\ket{\Psi_0} \) are the same many-body state. This degree of freedom has previously been exploited for various purposes, including the generation of localized molecular orbitals\cite{Boys1960, Edmiston1963, Pipek1989} and construction of an active-space in density matrix embedding theory(DMET)\cite{Knizia2012,Wouters2016}.
				
				If one wishes to use a basis of configurations to extract excited states, one must expect that the optimal and most efficient set of orbitals should capture the physical character (e.g. charge character) of the excited states (as opposed to the canonical orbitals).
				For a molecule near a metal surface, some of the most interesting nonadiabatic effects originate from the possibility of charge transfer between the molecule and the metal. This predicament indicates that, for our purposes, one would like to find a new set of orbitals which separate the molecule and metal components (even if there is strong mixing via  covalent bonds). 
				
				Here, we suggest rotating the canonical orbitals according to the following procedure. First, we project the localized molecular orbital onto the occupied and virtual spaces respectively:
				\begin{subequations}\label{eq:schmidt}
					\begin{align}
						\ket{d_o} &\equiv \frac{1}{\sqrt{\ev{n}}} \sum_i^{occ} \ket{i} \ip{i}{d} \\
						\ket{d_v} &\equiv  \frac{1}{\sqrt{1-\ev{n}}}\sum_a^{vir} \ket{a} \ip{a}{d}
					\end{align}
				\end{subequations}
				where \(\ev{n} = \sum_{i}^{occ} \abs{\ip{i}{d}}^2 \) is the impurity population of the ground state. Next, we orthonormalize the occupied and virtual spaces respectively while keeping \(\ket{d_o} \) and \(\ket{d_v} \) unchanged (this can be achieved by a QR decomposition with \( \ket{d_o}\) or \( \ket{d_v}\) being the first orbital). This leads to an occupied subspace space \(\{\ket{d_o}, \{\ket{\tilde{i}} \} \} \) and a virtual subspace \(\{\ket{d_v}, \{\ket{\tilde{a}} \} \} \). We refer to \(\{\ket{\tilde{i}} \} \)  as the occupied bath space, and \(\{\ket{\tilde{a}} \} \) the virtual bath space. Note that the bath orbitals \(\ket{\tilde{i}} \) and \(\ket{\tilde{a}} \) are not uniquely determined at this stage, because a unitary transformation within each bath space is still allowed. Finally, we demand that the bath orbitals diagonalize the Hamiltonian (Eq. \ref{eq:H:QIM}) in the bath space, namely,
				\begin{subequations}\label{eq:bath}
					\begin{align}
						\mel{\tilde{i}}{H}{\tilde{j}} &= \lambda_{\ti} \delta_{\ti\tj} \\
						\mel{\tilde{a}}{H}{\tilde{b}} &= \lambda_{\ta} \delta_{\ta\tb}
					\end{align}
				\end{subequations}
				which uniquely defines the orbital rotation.
				To summarize, we suggest an orbital rotation in each of the occupied and virtual subspaces in such a way that the Hamiltonian in the basis of occupied/virtual canonical orbitals now appears as follows
				\begin{equation}
					H_{occ/vir} = \left[
						\begin{array}{c|ccc}
							* & * & * & \dots \\\hline
							* & * & & \\
							* & & *&  \\
							\vdots & & & \ddots 
						\end{array}
					\right]
				\end{equation}
				This orbital rotation can be understood as a Schmidt decomposition and is similar to the DMET active space construction\cite{Knizia2012,Wouters2016}. In the context of DMET, \(\ket{d_o} \) is the Schmidt orbital related to \(\ket{d} \). 
				After the rotation, only \(\ket{d_o} \) and \(\ket{d_v} \) can possibly have non-zero impurity components; all the bath orbitals lie entirely in the metal.
			} 
			
			\subsubsection{Configuration Basis} {
				With the rotated orbitals defined above, configurations can be categorized according to whether they involve molecular excitations or not. For example, \( \ket{\Psi_{d_o}^{\tilde{a}}}\) represents a charge-transfer excitation from the molecule to the metal, while \(\ket{\Psi_{\tilde{i}}^{\tilde{a}}} \) is an EHP excitation located completely in the metal. For the purpose of simulating molecular nonadiabatic dynamics near a metal surface, configurations that involve molecular excitations are more important than pure-bath excitations; this idea lies at the heart of our reduced representation.
				
				We now make our first major approximation. We assume that only configuration interaction singles are needed for an accurate enough electronic structure;
				double excitations and beyond are completely ignored. This approximation can be understood as assuming ``fast bath equilibration", i.e., all configurations which involve more than two bath orbitals relax immediately. 
				For the pair of Schmidt orbitals, 
				note that \( \sqrt{\ev{n}} \ket{d_o} + \sqrt{1-\ev{n}} \ket{d_v} = \ket{d}  \) and its complement \(\ket{\bar{d}} \equiv -\sqrt{1-\ev{n}}\ket{d_o} + \sqrt{\ev{n}}\ket{d_v} \) is completely localized to the metal (\(\ip{d}{\bar{d}}=0 \) ), so \(\ket{d_o} \) and \(\ket{d_v} \) together contribute to one bath orbital. Specifically, if \(\epsilon_d \) is far below the Fermi level, then \(\ket{d_o} \approx \ket{d} \) and \(\ket{d_v}\) is almost completely localized to the metal; if \( \epsilon_d \) is far above the Fermi level, then \(\ket{d_v}\approx \ket{d} \) and \(\ket{d_o} \) is almost completely localized to the metal; if \(\epsilon_d \) is around the Fermi level, then both \(\ket{d_o} \) and \(\ket{d_v} \) are partially localized to the metal.
				
				The single excitation states contain the following four categories
				\begin{align}\label{}
					\{\ket{\Psi_{d_o}^{d_v}}, \{\ket{\Psi_{d_o}^{\tilde{a}}   } \}, \{\ket{\Psi_{\tilde{i}}^{d_v}} \}  , \{\ket{\Psi_{\tilde{i}}^{\tilde{a}} } \}     \} 
				\end{align}
				Among these four categories, the first three are relevant to the excitations on the molecule, while the last one merely contains bath excitations. We therefore define
				\begin{align}\label{}
					\Omega_{S} \equiv \{\ket{\Psi_0}, \ket{\Psi_{d_o}^{d_v}}, \{\ket{\Psi_{d_o}^{\tilde{a}}   } \}, \{\ket{\Psi_{\tilde{i}}^{d_v}} \}       \} 
				\end{align}
				to be the ``reduced system space'' of dynamical interest, and
				\begin{align}\label{}
					\Omega_{B} \equiv \{\ket{\Psi_{\tilde{i}}^{\tilde{a}} } \}     
				\end{align}
				to be the corresponding ``reduced bath space''. 
				
				We emphasize that \(\Omega_S \) and \(\Omega_B \) are not the physical impurity system or bath; states in \( \Omega_S\) have non-zero amplitude on the metal, and states in \(\Omega_B\) have non-zero amplitude on the molecule. 
				Here, \(\Omega_S \) contains a subset of the possible states of the entire system during a dynamical process. This subset is selected with a special focus on the molecular charge character, which we believe to be the most important ingredient for the molecular charge-transfer nondiabatic dynamics. Finally, to finish the construction of the reduced representation, let us now focus on \(\Omega_S \) and \(\Omega_B \) in turn.
				
				\begin{itemize}
					\item The reduced system (adiabatic) states, denoted \(\ket{\Psi_J}\), are defined as those which diagonalize the system Hamiltonian in \(\Omega_S \):
					\begin{subequations}\label{eq:system}
						\begin{align}
							&\ket{\Psi_{J}} = c_0^{(J)} \ket{\Psi_{d_o}^{d_v}} +  \sum_{\ti} c_{\ti}^{(J)} \ket{\Psi_{\tilde{i}}^{d_v}} + \sum_{\ta} c_{\ta}^{(J)} \ket{\Psi_{d_o}^{\tilde{a}}   } ~~~~(J>0)\\
							&\mel{\Psi_J}{H}{\Psi_{J'}} = E_J \delta_{JJ'}
						\end{align}
					\end{subequations}
					
					Note that the original ground state \(\ket{\Psi_0} \) does not couple to the single excitations, so it remains the ground adiabatic state in our reduced system; our reduced system has really just introduced a new set of excited adiabatic states. 
					
					\item \(\ket{\Psi_{\tilde{i}}^{\tilde{a}} } \) are defined to be the reduced bath states. By enforcing Eq. \ref{eq:bath}, these bath states are naturally the eigenstates of \(H\) in \(\Omega_B \):
					\begin{align}\label{}
						\mel{\Psi_{\tilde{i}}^{\tilde{a}}}{H}{\Psi_{\tilde{j}}^{\tilde{b}}} = \delta_{\ti\tj}\delta_{\ta\tb}\left(E_0 - \lambda_{\ti} + {\lambda}_{\ta} \right)
					\end{align}
				\end{itemize}
				The entire reduced representation contains the reduced system states and the reduced bath states as pictured schematically in Fig. \ref{fig:redrep}.

				The reduced representation defined above forms a new impurity model. 
				For the \(J-\)th system adiabatic state, we can define the hybridization function in a manner similar to Eq. \ref{eq:hybrid}:
				\begin{align}\label{eq:Gamma:rlx}
					\tilde{\Gamma}_{J}(E) = 2\pi \sum_{\tilde{i}\tilde{a} } \abs{\mel{\Psi_J}{H}{\Psi_{\tilde{i}}^{\tilde{a}} } }^2 \delta(E - E_{\tilde{i} }^{\tilde{a}}  ) ~~~~(J>0)
				\end{align}
				
				\begin{figure}
					\begin{subfigure}{.45\textwidth}
						\includegraphics[width=3in]{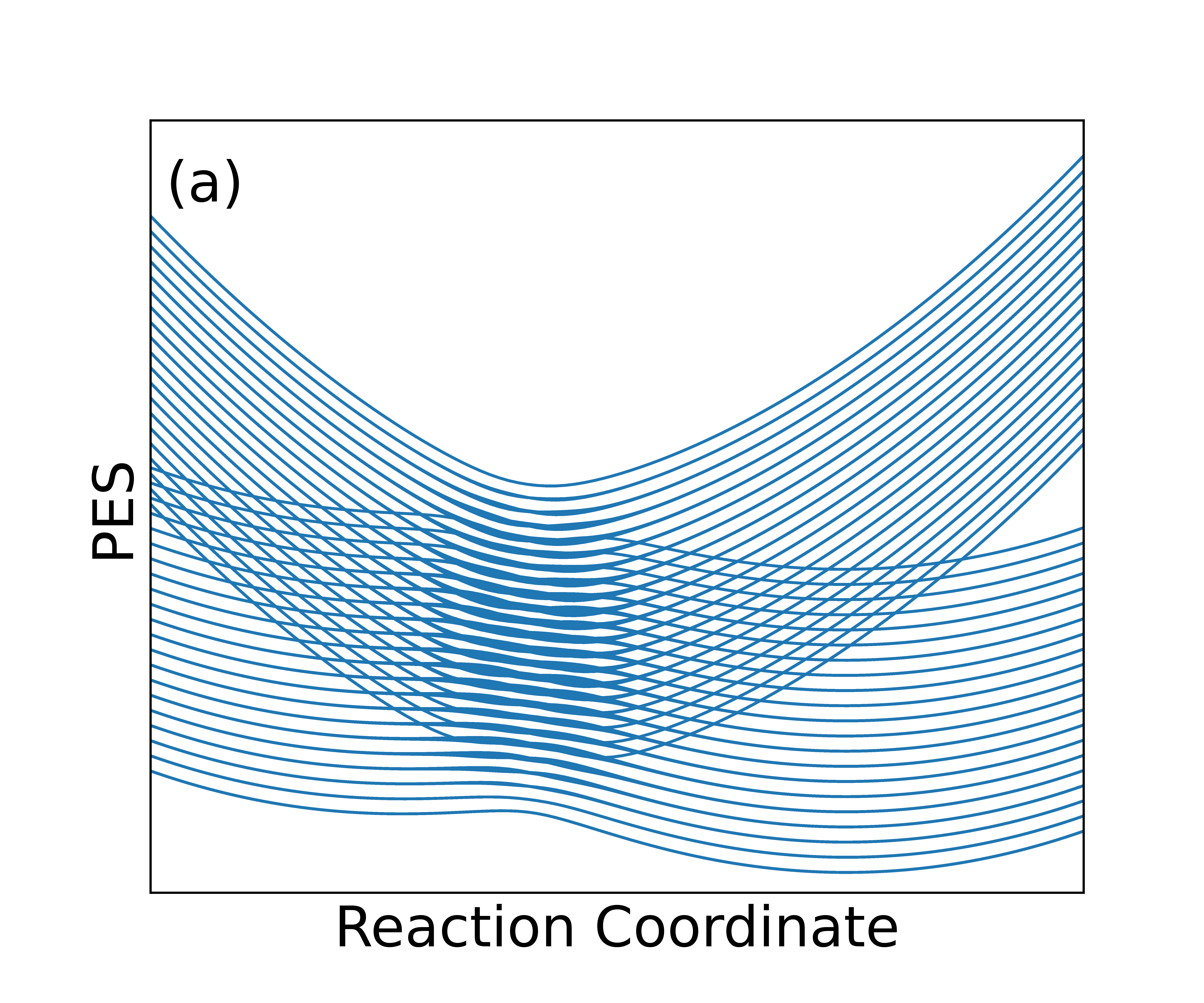}
					\end{subfigure}
					\begin{subfigure}{.45\textwidth}
						\includegraphics[width=2.75in]{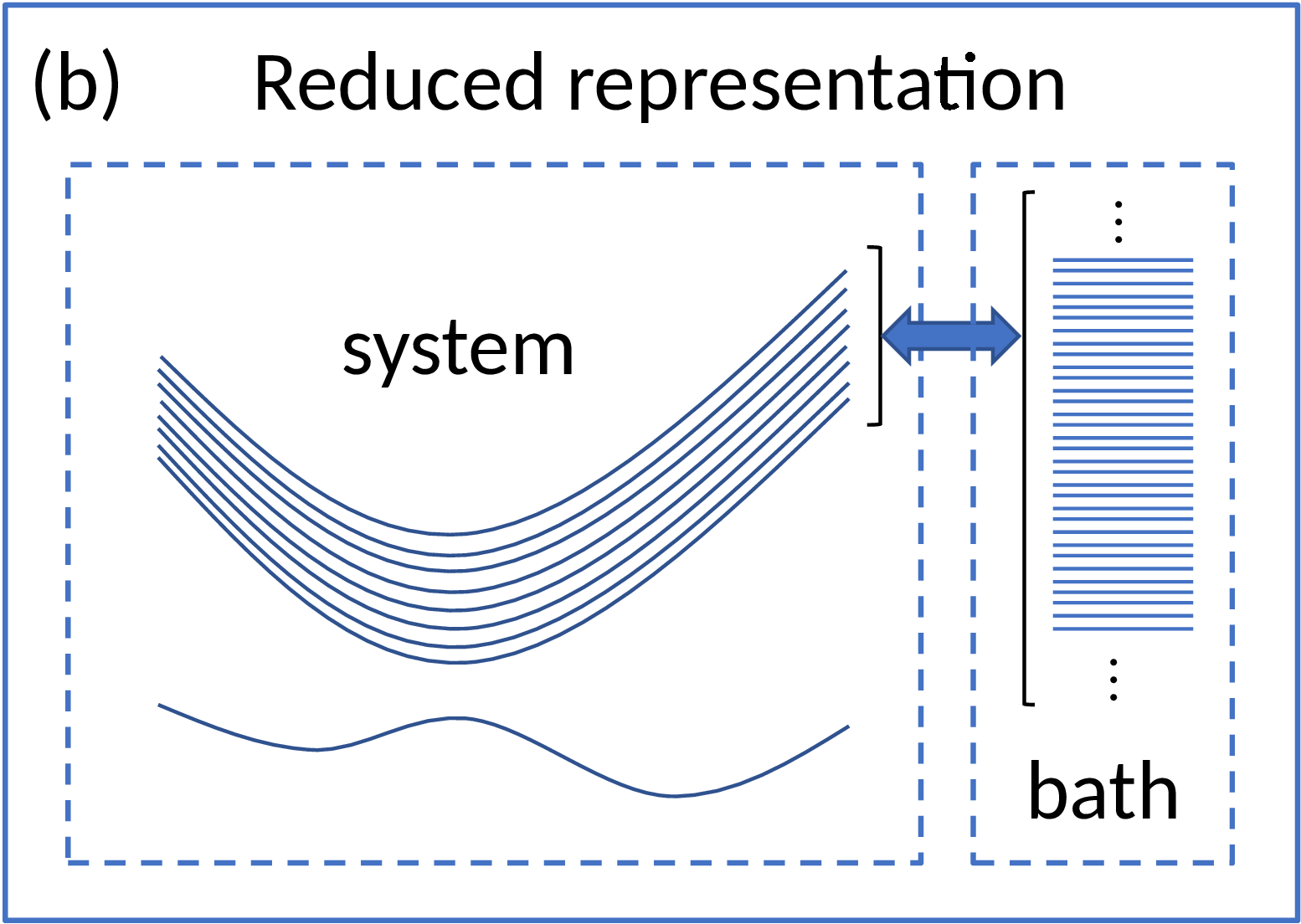}
					\end{subfigure}
					\caption{A schematic diagram of the reduced representation. (a) The original potential energy surfaces for a molecule-metal system. (b) A reduced representation for (a). Each excited state in the reduced system is coupled to the reduced bath.}
					\label{fig:redrep}
				\end{figure}
			} 
			
		} 
		
		\subsection{Fewest Switches Surface Hopping with Electronic Relaxation (FSSH-ER)} {
			\subsubsection{Electronic Dynamics} {
				Within conventional FSSH, all electronic states are evolved according to the full electronic Hamiltonian. 
				Armed with a finite number of adiabatic states and bath states defined above, we will now present a new variant of the FSSH method which includes explicit electronic relaxation to account for the presence of a truly infinite metal. 
				To do so, for a set of system states (defined in Eq. \ref{eq:system} and which are explicitly coupled to a bath), the system's electronic dynamics  follow a Lindblad equation in the Markovian limit:
				\begin{subequations}\label{eq:lindblad}
					\begin{align}
						\dv{\rho_S}{t} &= -\frac{i}{\hbar}[H_S, \rho_S] + \mathcal{L}(\rho_S) \\
						\mathcal{L}(\rho) &\equiv 
						\sum_{J} {\gamma}_J\left( L_J \rho L_J^\dagger - \frac{1}{2} \{ L_J^{\dagger} L_J, \rho \}  \right)
					\end{align}
				\end{subequations}
				Here \(\rho_S \) is the system density matrix, \(H_S \) is the Hamiltonian in \(\Omega_S \) subspace, \({\gamma}_J \) is a non-negative number related to the rate of relaxation, and \(L_J \) is the Lindblad jump operator.
				
				We now make our second major approximation. We assume that all of the system excited states \(\{\ket{\Psi_J}\} \) relax to the ground state at a rate given by the 
				new hybridization function defined in Eq. \ref{eq:Gamma:rlx} as evaluated at the relevant adiabatic energy, namely,
				\begin{align}\label{}
					\gamma_J = \tilde{\Gamma}_J(E_J)
				\end{align}
				This approximation is consistent with the premise of ``fast bath equilibration'', only now stronger; we assume that, once the system states relax to the bath states, those bath states immediately return to the ground state.
				
				As for the Lindblad jump operators, we choose them to be of the form
				\begin{align}\label{}
					(L_J)_{KL} = \delta_{K0}\delta_{LJ} \sqrt{1-f_J} + \delta_{KJ} \delta_{L0} \sqrt{f_J}~~~~~~~~(J > 0)
				\end{align}
				where
				\begin{align}\label{}
					f_J  \equiv \frac{1}{e^{\beta (E_J-E_0)} + 1}
				\end{align}
				It is easy to verify that the diagonal elements are
				\begin{subequations}\label{eq:Lrho:diag}
					\begin{align}\label{}
						\left(\mathcal{L}(\rho) \right)_{00} &= \sum_{J>0} \tilde{\Gamma}_J\left( \rho_{JJ}(1-f_J) - \rho_{00}f_J \right) \\
						\left(\mathcal{L}(\rho) \right)_{KK} &= \tilde{\Gamma}_K \left( \rho_{00} f_{K} - \rho_{KK}(1-f_K) \right) ~~~~(K>0)
					\end{align}
				\end{subequations}
				which ensure the correct electronic relaxation. The off-diagonal elements are 
				\begin{align}\label{}
					(\mathcal{L}(\rho))_{0K} &= \Gamma_K \sqrt{f_K(1-f_K)}\rho_{K0} - \frac{1}{2} \left( \Gamma_K(1-f_K)  + \sum_{J>0}\Gamma_J f_J    \right) \rho_{0K} ~~~~~~(K>0) \label{eq:lindblad:p0}\\
					(\mathcal{L}(\rho))_{JK} &= -\frac{1}{2} \left(\Gamma_J(1-f_J) + \Gamma_K(1-f_K) \right) \rho_{JK} ~~~~~~(J,K>0,~J\neq K)\label{eq:lindblad:pneqq}\\
					(\mathcal{L}(\rho))_{K0} &= (\mathcal{L}(\rho))_{0K}^*
				\end{align}
				In the zero-temperature limit (\(f_J=0\) for any \(J>0\)), Eqs. \ref{eq:lindblad:p0} and \ref{eq:lindblad:pneqq} reduce to
				\begin{align}\label{}
					(\mathcal{L}(\rho))_{0K} &= -\frac{1}{2} \Gamma_K \rho_{0K} ~~~~~~(K>0)\\
					(\mathcal{L}(\rho))_{JK} &= -\frac{1}{2} (\Gamma_J + \Gamma_K) \rho_{JK} ~~~~~~(J,K>0,~J\neq K)
				\end{align}
				
			} 
			\subsubsection{Surface Switching} {
				According to Tully's FSSH protocol, one switches between surfaces at a rate which guarantees that the proportion of nuclear trajectories on each potential energy surface agrees with the instantaneous density matrix. 
				To understand how this is achieved, note that, according to the time-dependent Schr\"odinger equation,
				\begin{align}\label{eq:drho:mm}
					\dv{\rho_{JJ}}{t} = -\sum_{K\neq J} 2\Re{T_{JK} \rho_{KJ}} \equiv -\sum_{K\neq J} g_{KJ} 
				\end{align}
				where \(T_{JK} \equiv \mel{J}{(\partial/\partial t)}{{K}} \) is the time-derivative coupling. The FSSH algorithm then requires that, at each time step, for a trajectory moving along state \(J\), one switches to state \(K\) in FSSH with probability
				\begin{align}\label{eq:hopprob}
					P_{K \leftarrow J} = \Delta t \frac{g_{KJ}}{\rho_{JJ}} \Theta(g_{KJ} )
				\end{align}
				where \(\Theta \) is the Heaviside step function. 
				If the fraction of nuclear trajectories agrees with the electronic density matrix at one time step, then, with the hopping probability above, the change in the population for state \(J\) is
				\begin{align}\label{}
					\Delta F_{J} &= \sum_{K\neq J} \left( - P_{K\leftarrow J} \rho_{JJ} +   P_{J \leftarrow K} \rho_{KK} \right) = \Delta t \sum_{K\neq J} \left(  -g_{KJ} \Theta(g_{KJ}) + g_{JK} \Theta(g_{JK})  \right)  \nonumber \\
					&= -\Delta t\sum_{K\neq J}g_{KJ} = \Delta \rho_{JJ}
				\end{align}
				Here we have used both Eq. \ref{eq:drho:mm} and the fact that \(g\) is anti-symmetric, \(g_{JK} = -g_{KJ} \). As such, after a time step \(\Delta t \), Tully's algorithm should keep the fraction of nuclear trajectories on each state in agreement with the electronic density matrix\cite{footnote:AFSSH}.
				
				With this consistency in mind, because of the extra relaxation term for  the electronic dynamics in Eq. \ref{eq:lindblad}, we will need to alter the surface switching algorithm accordingly. To be specific, we need to express Eq. \ref{eq:Lrho:diag} as a sum of anti-symmetric terms (just as in Eq. \ref{eq:drho:mm}). A convenient choice is
				\begin{align}\label{}
					\zeta_{KJ} &\equiv \delta_{K0} \tilde{\Gamma}_J \left(\rho_{JJ}(1-f_J) - \rho_{00} f_J \right) - \delta_{J0} \tilde{\Gamma}_K \left(\rho_{KK}(1-f_K) - \rho_{00} f_K \right)\label{eq:q}\\
					\dv{\rho_{JJ}}{t} &= -\sum_{K\neq J} \left(g_{KJ} + \zeta_{KJ} \right) \label{eq:gq}
				\end{align}
				Consequently, we also modify Eq. \ref{eq:hopprob} as follows:
				\begin{align}\label{}
					P_{K\leftarrow J} = \Delta t \frac{g_{KJ}+\zeta_{KJ}}{\rho_{JJ}} \Theta(g_{KJ}+\zeta_{KJ})
				\end{align}
				
				 
			} 
			
			\subsubsection{Momentum Rescaling and Velocity Reversal} {
				
				According to FSSH-ER, as one can see from Eq. \ref{eq:gq}, a hop can be initiated by either the derivative coupling (the \(g\) term) or electronic relaxation (the \(\zeta\) term). This state of affairs is to be contrasted with standard FSSH, where there is no electronic relaxation, which leads to another question vis-a-vis FSSH-ER. Namely,
				within standard FSSH, when a hop is successful, the momentum of the nuclear trajectory is rescaled to conserve the energy. This procedure is natural for a closed system, but is not appropriate for the open system we defined in Sec. \ref{sec:redrep}: after all, energy released by electronic relaxation is dissipated into a bath rather than the molecule, and the reduced system should not conserve energy. Therefore, consistent momentum rescaling at every hop is not appropriate, and a new protocol is required for FSSH-ER.
				
				To address this issue, we suggest that, if a successful hop is initiated by the derivative coupling, the momentum is rescaled as usual; if a hop is initiated by the electronic relaxation, the momentum is not rescaled. Specifically, consider a successful hop from state \(J\) to state \(K\). If \(g_{KJ}>0 \) and \(\zeta_{KJ}<0\), momentum is rescaled as usual; if \(g_{KJ}<0 \) and \(\zeta_{KJ}>0\), momentum is not rescaled; if \(g_{KJ}>0\) and \(\zeta_{KJ}>0\), an additional random number \(r\) with \(0<r<1 \) is generated and the momentum is rescaled if \(r < g_{KJ}/(g_{KJ}+\zeta_{KJ}) \). 
				
				Now, in practice, another important component for FSSH is the velocity reversal as suggested by Japser and Truhlar\cite{Jasper2003}, which has been shown to be a necessary ingredient for rate simulations\cite{Jain2015,Jain2015-2}. Just as above, we recommend that this velocity reversal be invoked only if a frustrated hop is initiated by the derivative coupling.
			} 
			
		} 
	} 

	\section{Results} \label{sec:result} {
		To test the validity of our method, we will make the standard, simple approximation of a pair of parabolic diabatic PESs
		\begin{subequations} \label{eq:pes:diab}
			\begin{align}\label{}
				U_0(x) &= \frac{1}{2} m\omega^2 x^2 \\
				U_1(x) &= U_0(x) + \epsilon_d(x) = \frac{1}{2} m\omega^2 (x-x_1)^2 + \Delta 
			\end{align}
		\end{subequations}
		We will discuss more realistic Hamiltonians in Sec. \ref{sec:discussion}. For the present, parabolic Anderson-Holstein model, a great deal is known about the relevant nonadiabatic dynamics.
		In the \(kT \gg \hbar\omega \gg \Gamma\) limit, the electron transfer rate can be calculated by the Fermi's Golden Rule and reduces to the Marcus theory of electrochemical electron transfer:
		\begin{subequations}\label{eq:marcus}
			\begin{align}\label{}
				k_{Marcus}&= k_{0\rightarrow 1} + k_{1\rightarrow 0} \\
				k_{0\rightarrow 1} &= \int_{-\infty}^{+\infty} \dd \epsilon f(\epsilon)  \frac{\Gamma(\epsilon)}{\hbar} \sqrt{\frac{1}{4 \pi E_r kT  }} \exp(   -\frac{( E_r + \Delta -\epsilon  )^2}{4E_r kT }   )\\
				k_{1\rightarrow 0} &= \int_{-\infty}^{+\infty} \dd \epsilon (1-f(\epsilon))  \frac{\Gamma(\epsilon)}{\hbar} \sqrt{\frac{1}{4 \pi E_r kT  }} \exp(   -\frac{( E_r - \Delta +\epsilon  )^2}{4E_r kT }  )
			\end{align}
		\end{subequations}
		Here, \(E_r\equiv m\omega^2 x_1^2/2 \) is the reorganization energy and \(\Gamma(\epsilon) \equiv 2\pi \sum_k \abs{V_k}^2 \delta(\epsilon-\epsilon_k) \).
		In the \(kT,\Gamma\gg \hbar \omega \) limit, the rate is given by transition state theory
		\begin{align}\label{eq:tst}
			k_{TST} = \kappa \frac{\omega}{2\pi} e^{-U^{(b)}/kT}
		\end{align}
		where \(U^{(b)} \) is the barrier height and \(\kappa \) is the transmission coefficient.
		
		For simplicity, the system is taken to be in the wide-band limit. In other words, we assume that (i) the set of energies \( \{\epsilon_k \}\) spans a sufficiently wide range and (ii) the hybridization function  is independent of energy. 
		The parameters used in our simulation are \(m=2000 \), \(\omega=0.0002 \), \(x_1 = 20.6097\), \(\Delta=-0.0038 \). The metal states \(\{\epsilon_k \} \) span from \(-0.2 \) to \(0.2\) and are evenly spaced. The number of metal states is chosen to converge the PESs and derivative couplings, as will be discussed below. The temperature in our simulation is chosen to be \(kT=9.5\times 10^{-4}\).
		
		\subsection{PESs and Derivative Couplings}\label{sec:result:pesdc} {
			In Fig. \ref{fig:pesdc}, we plot the adiabatic PESs and the derivative couplings between the ground state and the excited states, with (a) \(\Gamma=0.0064 \) and (b) \(\Gamma = 0.0008 \). The energy spacing between metal orbitals (\(\delta \epsilon_k\)) is \(1\times 10^{-3}\) and \(1.25\times 10^{-4} \) respectively, which is about \(\Gamma/6.4 \).
			In both cases, the ground PESs have the shape of a  double-well, and the lowest excited PESs recover the diabatic PESs asymptotically. 
			The derivative coupling between the ground state and the first excited state peaks around the diabatic crossing point and is nearly zero elsewhere.
			As \(\Gamma\) becomes smaller, all the PESs approach the diabatic PESs, and the derivative couplings grow but narrow at the diabatic crossing point. 
			Thus, our calculations seem analogous to electronic structure in solution with  \(\Gamma \) playing the role of the diabatic coupling \(V\).
			And yet this analogy cannot be strictly correct, given that \(V\) applies when there are two electronic states, and \(\Gamma\sim 2\pi V^2\rho \) applies when there is a continuous density of states.
			
			To better understand Fig. \ref{fig:pesdc}, note that, for the full Hamiltonian in Eq. \ref{eq:H:QIM}, one expects that the lowest possible excitation energy will be roughly the energy spacing between metal orbitals, \(\delta \epsilon_k \). 
			After all, low-lying metal excitations are always possible. 
			And so, as \(\delta \epsilon_k \) approaches zero, so should the excitation gap. 
			Here, however, we find that the excitation gap near the diabatic crossing converges to a finite value close to \(\Gamma\) as shown in Fig. \ref{fig:gap}. 
			Apparently, by choosing a CIS subspace to reflect charge-transfer excitations alone, we have successfully excluded pure-bath excitations above the ground state (with energies lower than the lowest charge-transfer excitation).
			In doing so, 
			we have dramatically reduced the computational cost of the FSSH-ER approach, allowing one to focus on the adiabatic states with large derivative couplings to the ground state.
			Moreover, as shown in Fig. \ref{fig:dcmax}, the derivative couplings between high-energy states with the ground state do become small. Thus, we expect that 
			a simulation of charge-transfer dynamics in our reconstructed system can (at least sometimes) be performed with merely a handful of PESs. 
			Finally, for a rough explanation of why the energy gap between the ground state and excited states seemingly approaches the value of \(\Gamma\) per se, please see Appendix \ref{apdx:gap}.
			
		} 
		
		\subsection{Relaxation \(\tilde{\Gamma}(E_J)\) }\label{sec:Gamma:rlx} {
			One major difference between our method and conventional FSSH or IESH is the presence of explicit electronic relaxation as characterized by \(\tilde{\Gamma}_J(E_J)\) and defined in Eq. \ref{eq:Gamma:rlx}.
			In Fig. \ref{fig:gamma}(a), we plot \(\tilde{\Gamma}_1 \) (the relaxation for the first excited state) as a function of nuclear coordinates at four different hybridization \(\Gamma\)'s. 
			We find that, when the nuclear coordinates are far away from the diabatic crossing,  \(\tilde{\Gamma}_1 \approx \Gamma\) which agrees with our intuition of electronic relaxation (that is independent of nuclear motion). Interestingly, \(\tilde{\Gamma}_1 \) displays a dip at the diabatic crossing (where the derivative coupling is large), and the relative depth of this dip increases as \(\Gamma \) decreases. 
			This state of affairs gives us a satisfying view of nonadiabatic effects at a molecule-metal interface: at the crossing point, there is a large derivative coupling (to accommodate nuclei switching surfaces) and a small \(\tilde{\Gamma}\) (to accommodate electronic relaxation that is independent of nuclear motion); far from a crossing point, however, one finds a large \(\tilde{\Gamma}\) and a small derivative coupling.
			
			Next, we turn our attention to the behavior of  \(\tilde{\Gamma}_J\)  as a function of excitation state energy.
			In Fig. \ref{fig:gamma}(b), we plot \(\tilde{\Gamma}_J\) for \(J=1,2,5,10 \) at \(\Gamma=0.0008\). For higher-excited states, we continue to find that \(\tilde{\Gamma}_J \approx \Gamma\) far from the crossing. However, in the vicinity of the crossing, \(\tilde{\Gamma}_2\) and \(\tilde{\Gamma}_5\) have a bump rather than a dip, and \(\tilde{\Gamma}_{10}\) displays a curious oscillating pattern.
			Apparently, it is difficult to find an intuitive picture of nonadiabatic effects between many electronic states in the limit of a continuum: the Born-Oppenheimer formalism of generating adiabatic states is not directly compatible with a reduced description of charge transfer, and all of the complications created by the Born-Oppenheimer treatment lead to very intriguing
			behavior of the high-lyding excited state  \(\tilde{\Gamma}_{J}\) near the diabatic crossing. The form of these \(\tilde{\Gamma}_{J}\) functions will be investigated in a future publication.
			
		} 
	
		\subsection{Electron Transfer Rate} {
			Finally, in Fig. \ref{fig:rate}, we compare the electron transfer rate predicted by FSSH-ER as a function of \(\Gamma \) with three other methods: (1) Marcus theory, which is valid in the small \(\Gamma \) limit; (2) transition state theory(TST), which is valid in the large  \(\Gamma \) limit; (3) BCME, which interpolates between both limits. 
			These results have been previously reported in Ref. \citenum{Ouyang2016}.
			To test the FSSH-ER method above, we perform a simulation with all trajectories initialized as the ground state of the left well and subject to an external nuclear friction \(\gamma_{n} = 2m\omega\) (and the corresponding random force that obeys the fluctuation-dissipation theorem). 
			Each data point is obtained by averaging over 2400 classical trajectories on 30 PESs. The rate is obtained by fitting \(\ev{\dg{d}d} \) to the function \(Ae^{-kt} + B \) where \(k\) is the rate constant. 
			
			According to Fig.  \ref{fig:rate}, our method agrees with transition state theory (with \(\kappa=0.5\)) in the large \(\Gamma\) limit. In the small \(\Gamma\) limit, our method does predict a rate that decreases as \(\Gamma\) decreases, as does Marcus theory. However, for the smallest \(\Gamma\), our method differs from the Marcus rate by about a factor of 1.6, and the slope is also different. 
			Such differences are known for FSSH-like methods, especially without any decoherence\cite{Subotnik2011}, which usually lead to an overestimate of the rate in the small \(\Gamma\) limit. 
			
			Finally, in Fig. \ref{fig:rate}, we also plot  data  from a simulation which does not include any electronic relaxation, i.e., the electronic equation of motion obeys the quantum Liouville equation, and the surface switching algorithm merely considers the derivative couplings. In other words, we set \(\mathcal{L}=0 \) in Eq. \ref{eq:lindblad} and \(\zeta_{KJ}=0\) in Eq. \ref{eq:q}. 
			As Fig. \ref{fig:rate} shows conclusively, for small \(\Gamma\), electronic relaxation in our simulation is crucial: the predicted rates are significantly underestimated without electronic relaxation. 
			Thus, Fig. \ref{fig:rate} would appear to validate a new picture of electron transfer at a metal surface that can interpolate between the transition state theory limit (large \(\Gamma\), with broadening) and the Marcus limit (small \(\Gamma\)); one can effectively include both derivative couplings and explicit electronic relaxation at different points in coordinate space.	
			
		} 
		\begin{figure}
			\includegraphics[width=6.6in]{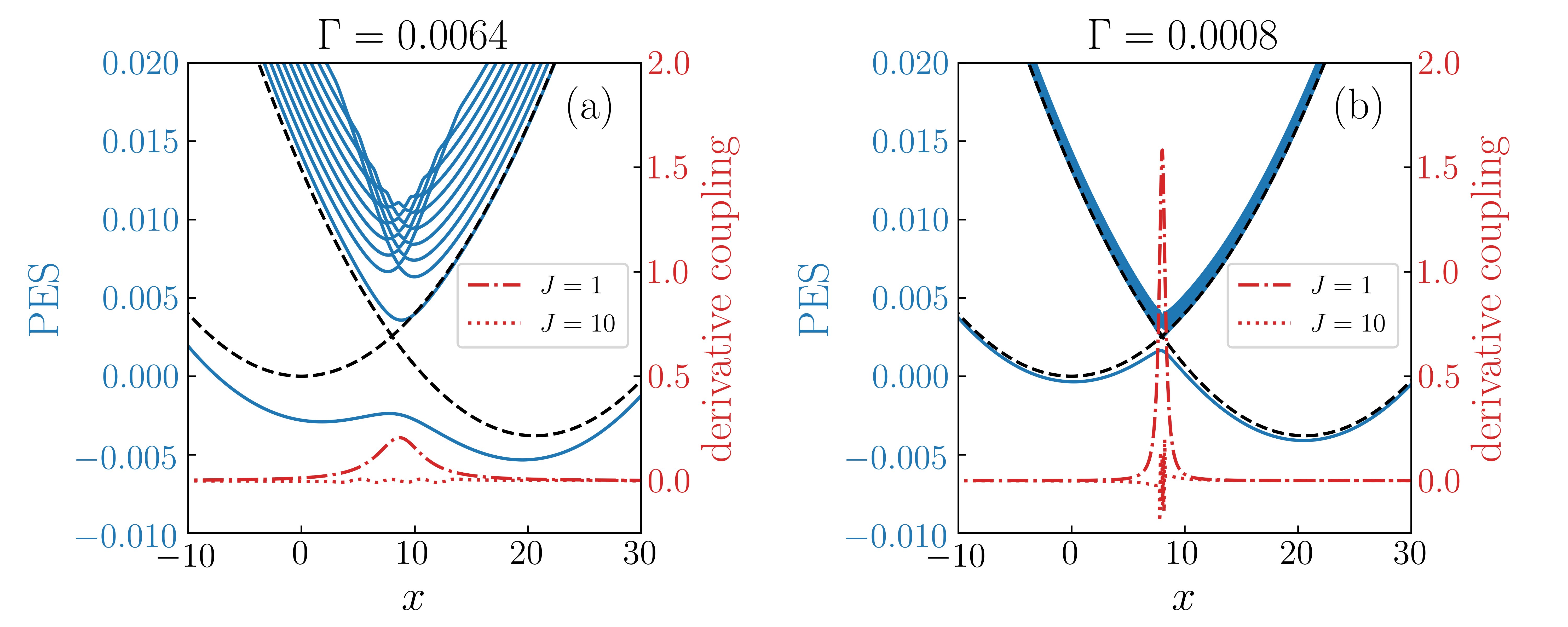}
			\caption{Subspace adiabatic potential energy surfaces and derivative couplings with (a)  \(\Gamma=0.0064\), and (b) \(\Gamma=0.0008\). In both figures, the one-electron metal states \(\{\epsilon_k \} \) are evenly spaced and range from -0.2 to 0.2 with a spacing of \(\Gamma/6.4\). The other parameters are \(m=2000 \), \(\omega=0.0002 \), \(x_0 = 0\), \(x_1 = 20.6097\), \(\Delta=-0.0038 \). The ground adiabatic PES and ten lowest excited PESs are plotted in solid lines. The diabatic PESs (Eq. \ref{eq:pes:diab}) are plotted in dashed lines for reference. The derivative coupling between the ground state and the first excited state peaks at the diabatic crossing. For higher excited states, the derivative couplings with the ground state are much smaller. Note that there is a finite gap between the ground and excited PES, indicating that pure-bath excitations of the ground state are excluded.}
			\label{fig:pesdc}
		\end{figure}
	
		\begin{figure}
			\includegraphics[width=3.3in]{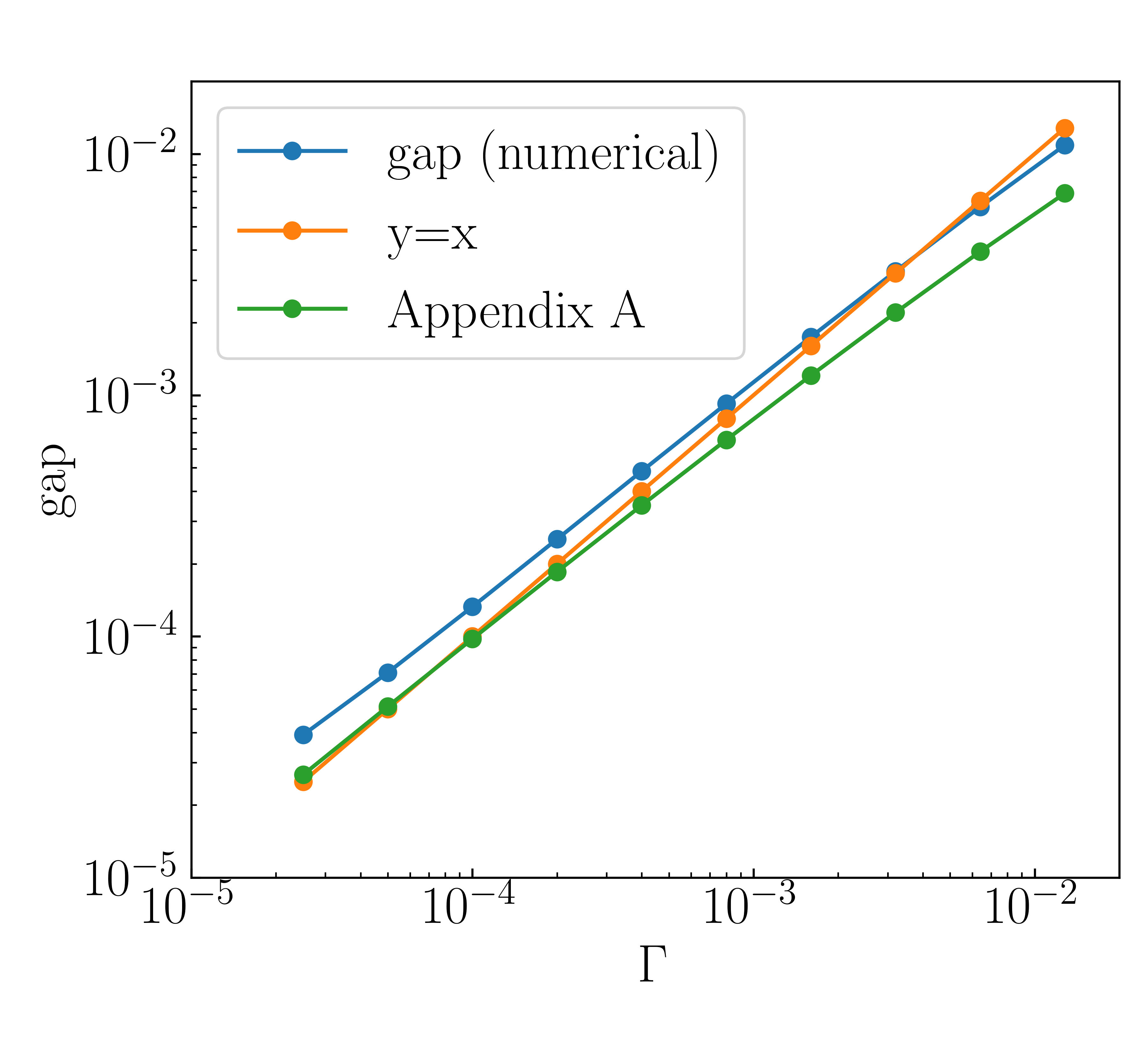}
			\caption{Excitation gap at the diabatic crossing vs. hybridization \(\Gamma \). A rough explanation for why the gap \(\approx \Gamma \) is given in Appendix \ref{apdx:gap}. }
			\label{fig:gap}
		\end{figure}
		
		\begin{figure}
			\includegraphics[width=6.6in]{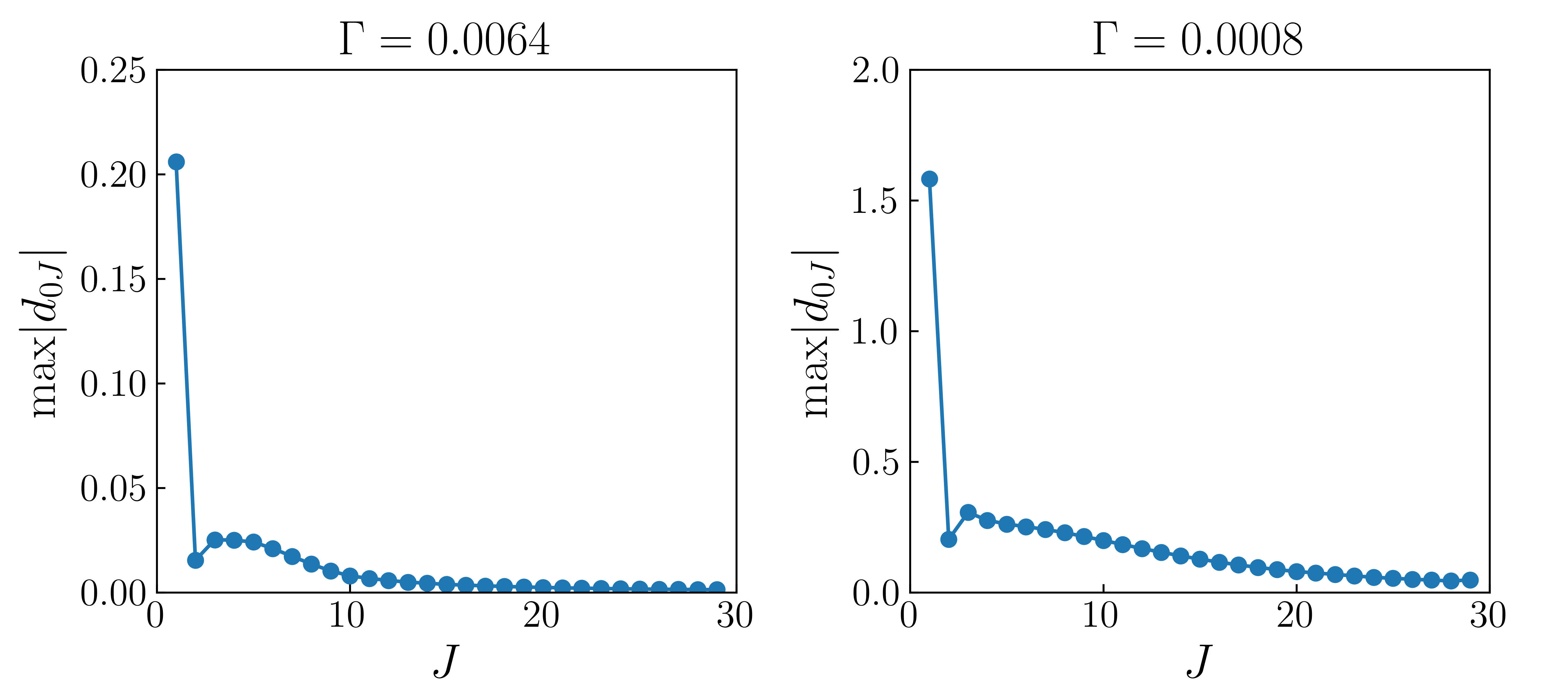}
			\caption{Maximum derivative couplings between the ground state and excited states. Note that the derivative couplings become small in all cases for higher excited states (and sometimes much faster).}
			\label{fig:dcmax}
		\end{figure}
	
		\begin{figure}
			\includegraphics[width=6.6in]{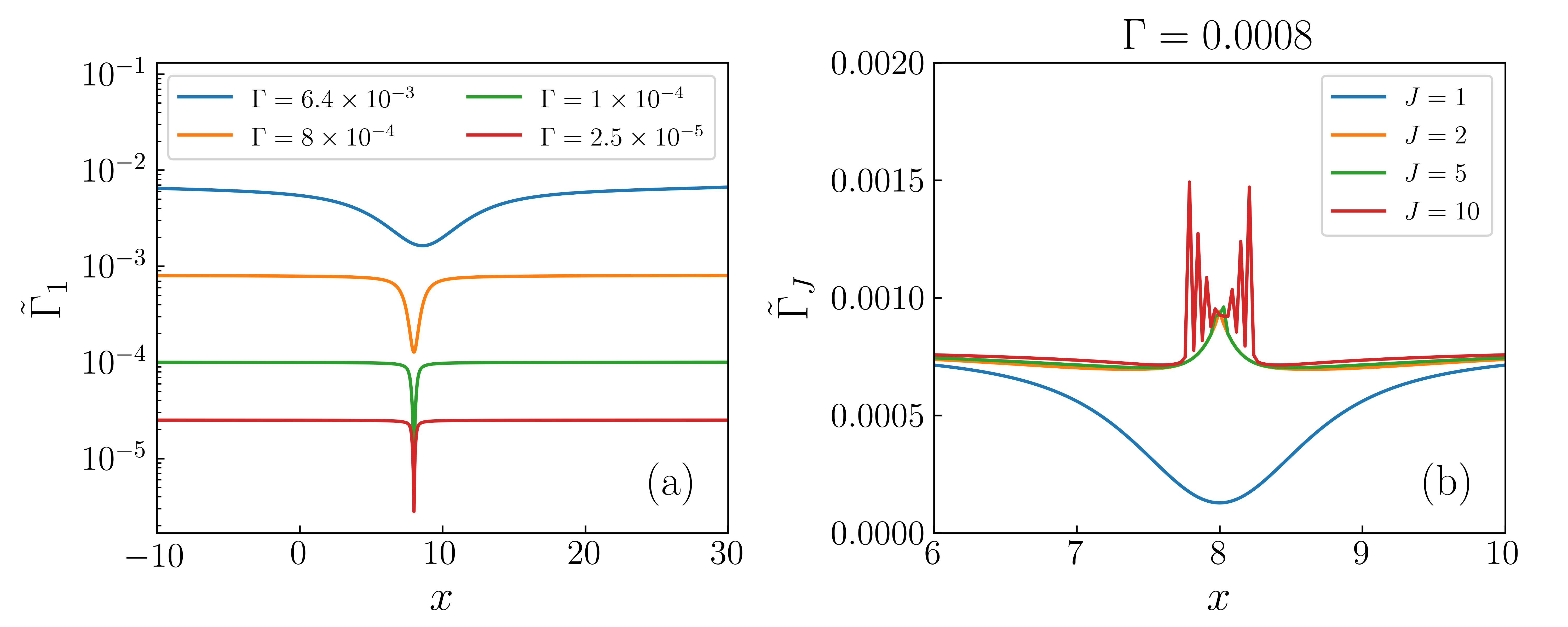}
			\caption{Relaxation \(\tilde{\Gamma} \) (Eq. \ref{eq:Gamma:rlx}) as a function of nuclear coordinate. The parameters are the same as those in Fig. \ref{fig:pesdc}. The diabatic PESs cross at \(x=8.0011\). (a) \(\tilde{\Gamma} \) for the first excited state at four different hybridization \(\Gamma\)'s. Note that, except for the vicinity of the crossing, we find \(\tilde{\Gamma}_1 \approx \Gamma\). For \(\tilde{\Gamma}_1 \), a ditch is always present near the crossing, highlighting the idea that nonadiabatic nuclear dynamics dominate  at the crossing point (where the derivative coupling is large), but standard electronic relaxation (whereby molecular electrons exchange with the metal) dominate everywhere else. (b) \(\tilde{\Gamma}_J\) for \(J=1,2,5,10 \) at \(\Gamma=0.0008\). Unlike \(\tilde{\Gamma}_1\), \(\tilde{\Gamma}_J\) for higher-excited states can exhibit a bump (\(J=2\) and 5) or even an intriguing oscillating pattern (\(J=10\)) near the crossing. }
			\label{fig:gamma}
		\end{figure}

		\begin{figure}
			\includegraphics[width=3.3in]{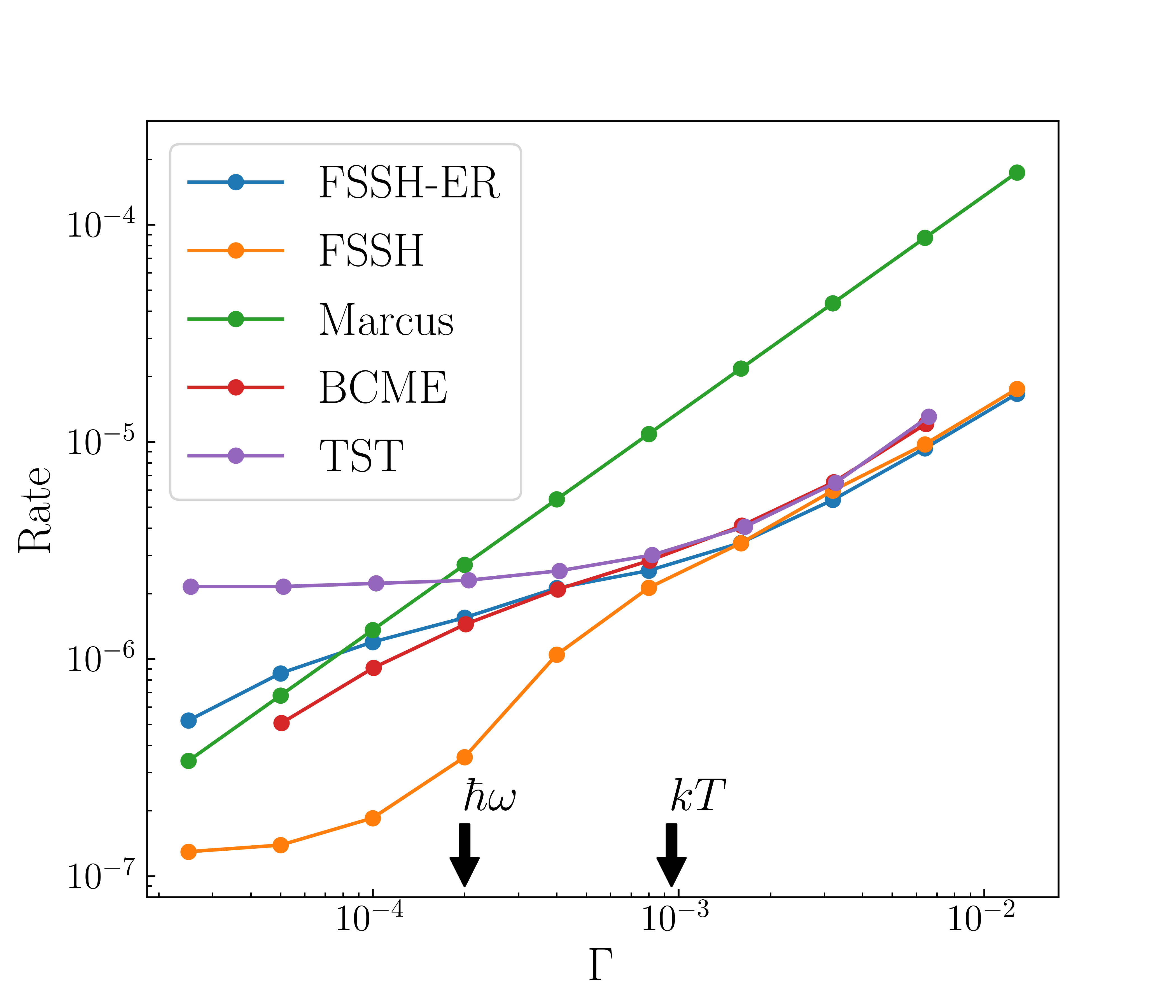}
			\caption{Rate of electron transfer as a function of \(\Gamma \). The parameters are the same as those in Fig. \ref{fig:pesdc}. Marcus theory is valid in the small \(\Gamma\) limit, and transition state theory (TST) is valid in the large \(\Gamma\) limit (\(\kappa=0.5\)). Overall both BCME and FSSH-ER are reliable across the full range of \(\Gamma\). In the small \(\Gamma\) limit, FSSH-ER differs slightly from the Marcus rate in part due to a lack of decoherence\cite{Subotnik2011}. For small \(\Gamma\), ignoring the electronic relaxation in the reduced system can significantly underestimate the rate.}
			\label{fig:rate}
		\end{figure}
	} 
	
	\section{Discussion}\label{sec:discussion} {
		Having demonstrated the power of the present FSSH-ER approach, let us now discuss several nuances of the approach as well as the future directions and possibilities.
		\subsection{Convergence Issues} {
			For a realistic system, a metal contains a continuum of electronic states. For practical simulations, however, this continuum is always replaced by a set of discretized states that form a ``quasi-continuum''. One immediate question is, what is the criterion for a good quasi-continuum?
			Here, in our system-bath reconstruction procedure, we notice that the excitation gap and derivative couplings between our subspace adiabatic PESs seemingly converge when the energy spacing of this quasi-continuum is smaller than the hybridization \(\Gamma\). This criterion poses a challenge for a system in the small \(\Gamma \) limit: for example, for a  realistic calculation, this criterion would demand an ultra-dense Brillouin zone sampling. One possible solution would to use Wannier interpolations\cite{Giustino2007}. For now, our intentions is to use FSSH-ER only when a molecule is reasonably close to a surface so that \(\Gamma\) should not become too small. 
			This circumstance is the most crucial case for electrochemical and catalytical simulations, as the limit \(\Gamma\rightarrow 0\) can usually be treated perturbatively with Marcus theory (or some variant thereof).
			
			Another question about convergence regards the number of PESs used in the dynamical simulation. For high-energy excited states, the derivative couplings (\(d_J\)) with the ground state becomes less significant, suggesting that there must be a natural cutoff. In this work, we used 30 PESs for our dynamical simulations based on the criterion \( \text{max}(\abs{d_J}) <  \text{max}(\abs{d_1})/20\). However, such a cutoff based on relative magnitudes might not be sufficient; a cutoff based on the absolute magnitude might be necessary. 
			Overall, assuming (1) there is no photo-excitation, (2) the system is initially thermally equilibrated and (3) is initiated on the ground state, the present algorithm appears robust. Otherwise, the importance, relevance and necessity of including many high-energy excited states will need to be addressed in the future.
		} 

		\subsection{Multiple Molecular Orbitals and Electron-Electron Interactions}\label{subsec:multi_ee} {
			In the present work, we consider only a non-interacting model where a molecule can be represented as a single impurity orbital. For realistic systems, such a model can hardly be adequate. Below, we will discuss two aspects that go beyond the model we considered above.
			
			First, for many chemical problems of interest, there can be multiple molecular orbitals which are energetically relevant to a charge-transfer process near the metal surface. Suppose there are \(N\) energetically relevant molecular orbitals. One may project each orbital onto the occupied and virtual spaces individually, yielding \(N\) Schmidt occupied and virtual orbitals respectively. Alternatively, if one uses a localized basis, one can perform a singular value decomposition to the molecular block of the occupied (virtual) orbitals and use the right-singular vectors to rotate the orbitals\cite{Knizia2012,Wouters2016}.
			Thus, it is very likely that the present approach should be extendable to the case of many molecular orbitals, albeit with a higher computational cost. And indeed, in Ref. \citenum{Chen2020}, we show how to generate relevant electronic states for a molecule composed of two molecular orbitals. However, in Ref. \citenum{Chen2020}, we address only the ground state and not excited states or dynamics. Extending and benchmarking the present nonadiabatic formalism to the case of many molecular orbitals is a crucial step forward for this research.
	
			Second, the inevitable elephant in the room when we model electron transfer at a molecule-metal interface is always the electron-electron interactions, which can scarcely be ignored in \textit{ab initio} simulations.
			For a molecule alone, the Hartree-Fock approximation often gives qualitatively wrong results in the presence of strong e-e interactions, and one usually must resort to a higher level of electronic structure methods. 
			Now obviously, for a molecule at a metal surface, Hartree-Fock is not an option, but DFT has proven to be very effective for modeling surface calculations. And since DFT takes the guise of an effective mean-field theory, the electronic structure and dynamics used above should be immediately applicable.  In this regard, merging the current FSSH-ER formalism with DFT will be a top priority for future research. 
			
			Of course, if bonds are broken and/or one works in the small Gamma limit, standard DFT may fail and need to be adjusted. Now, in a previously published study, we have shown that, for an isolated and twisted C2H4 molecule, one can improve upon DFT by including one double excitation\cite{Teh2019}. Moreover, in Ref. \citenum{Chen2020}, we have also shown that including a subset of double excitations can vastly improve the performance of an electronic structure method describing a molecule on a metal surface (at least as far as ground state properties).  In the future, it will be very exciting to merge FSSH-ER with {\em correlated} electronic structure techniques, for a truly robust view of nonadiabatic dynamics at a metal surface. This work is ongoing in our laboratory.
			
		}

	} 
	\section{Conclusion}\label{sec:conclusion} {
		In this article, we have described a new method for simulating the coupled electronic-nuclear dynamics of a molecule near a metal surface with a special focus on molecular charge-transfer nonadiabatic effects.
		Starting with a molecule-metal system's one-electron eigenstates, we build a set of configuration states where the impurity-related excitations can be distinguished from the pure-bath excitations. Next, a reduced representation is constructed for the purpose of dynamics. Then, based on this representation, we have proposed a modified surface hopping scheme with explicit electronic relaxation.
		Finally, this method has been tested in a non-interacting Anderson-Holstein model, and we have extracted electron transfer rates. Our results appear valid across the full range of \(\Gamma\).
		
		Although the present work is limited to a non-interacting system with one impurity orbital, the framework established here can be easily extended to \textit{ab initio} mean-field calculations with multiple impurity orbitals. We will also investigate multiple impurity orbitals with strong electron-electron interactions beyond mean-field theory in the future. 
		While practical questions do remain regarding how many states are required for convergence and the behavior of high-lying excited states in the reduced system, the FSSH-ER protocol appears to provide an efficient strategy for simulating molecular charge-transfer nonadiabatic dynamics both in the adiabatic and nonadiabatic regimes.
		Looking forward, our next step is to combine the present algorithm with \textit{ab initio} electronic structure methods and hopefully make contact with realistic problems in electrochemistry and heterogeneous catalysis.

	} 

	\begin{acknowledgement}
		This work was supported by the U.S. Air Force Office of Scientific Research (USAFOSR) AFOSR Grants No. FA9550-18-1-0497 and FA9550-18-1-0420.  Computational support was provided by
		the High Performance Computing Modernization Program (HPCMP) of the Department of Defense.
	\end{acknowledgement}

	\appendix
	\section{An Estimate of the Excitation Gap at the Crossing} \label{apdx:gap} {
		As mentioned in Sec. \ref{sec:result:pesdc}, by choosing a CIS subspace which excludes pure-bath excitations and focusing on charge-transfer excitations, we find a non-zero excitation gap in our redefined system.
		In this appendix, we will give a rough estimate of this gap near the diabatic crossing point.
		
		Given any non-interacting impurity Hamiltonian
		\begin{align}\label{eq:H}
			H = \begin{bmatrix}
			\epsilon_d & \cdots & {V}_k^* & \cdots \\
			\vdots & \ddots \\
			{V}_k & & \epsilon_k \\
			\vdots & & & \ddots
			\end{bmatrix}
		\end{align}
		the eigenvalues, denoted \(\lambda\), are given by the characteristic equation \(\det(\lambda I-H)=0 \), which is
		\begin{align}\label{eq:chareq:raw}
			\left(\lambda - \epsilon_d  \right) \prod_{k}\left(\lambda - \epsilon_k\right) - \sum_{k} {\abs{V_k}^2} \prod_{k'}^{k'\neq k} \left(\lambda - \epsilon_{k'} \right) = 0
		\end{align}
		Let us assume that bath states that do not couple to the impurity can be ignored, so that (1) \(V_k \neq 0\) and (2) \(\epsilon_k \neq \epsilon_{k'} \) for \(k \neq k' \). (For the second assumption, if  there are degenerate bath levels, we can always rotate them through a Householder reflection so that only one of them couples to the impurity). Therefore, Eq. \ref{eq:chareq:raw} is equivalent to
		\begin{align}\label{eq:chareq}
			\lambda - \epsilon_d - \sum_k \frac{\abs{V_k}^2}{\lambda-\epsilon_k} = 0
		\end{align}
		In other words, the eigenvalues of Eq. \ref{eq:H} are the roots of Eq. \ref{eq:chareq}.
		
		Now, the Hamiltonian for our reduced CIS subspace \(\{\ket{\Psi_{d_o}^{d_v}}, \ket{\Psi_{d_o}^{\tilde{a}}}, \ket{\Psi_{\tilde{i}}^{d_v}} \) is of the form given in Eq. \ref{eq:H} and reads
		\begin{align}\label{eq:exc}
			H_{CIS} &= (E_0 + \lambda_v-\lambda_o) I + \mqty[0  & \mel{d_v}{H}{\tb} & -\mel{\tj}{H}{d_o} \\
			\mel{\ta}{H}{d_v} &  \lambda_{\ta} \delta_{\ta\tb} - \lambda_v I  & 0 \\
			-\mel{d_o}{H}{\ti} & 0 &  \lambda_o I - \lambda_{\ti} \delta_{\ti\tj} ] \nonumber \\
			&\equiv (E_0 + \lambda_v-\lambda_o) I + \tilde{H}_{CIS}
		\end{align}
		Here, \(\lambda_{(\ldots)} \) represents the energy of a rotated orbital; for example, \(\lambda_o \equiv \ev{H}{d_o}  \). The problem of finding the relevant excitation gap is equivalent to finding the smallest eigenvalue (\(\lambda\)) of \(\tilde{H}_{CIS} \) at the diabatic crossing, where \(\ev{\dg{d}d}=0.5 \). We found numerically in Sec. \ref{sec:result:pesdc} that \(\lambda \approx \Gamma-\lambda_v+\lambda_o\) and we would like to confirm this result analytically.
		
		To make progress, we begin by estimating the orbital energy of the Schmidt orbital \(\ket{d_o}\) that is dual to the impurity. For simplicity, consider a band than spans from \( -W\) to \(W\) with a constant hybridization function \(\Gamma \). Assume that the Fermi level \(\epsilon_F=0 \), and the impurity on-site energy is \(\epsilon_d=0 \). Then,
		\begin{align}\label{eq:imppop}
			\lambda_o &= \frac{1}{\ev{n}} \sum_{i}^{occ} \lambda_i \abs{\ip{i}{d}}^2 = 2 \int_{-W}^{0}   \epsilon \frac{1}{\pi}\frac{\Gamma/2 }{(\epsilon - \Lambda(\epsilon))^2 + (\Gamma/2)^2  }  \dd \epsilon
		\end{align}
		where \(\Lambda(\epsilon) \) is the real part of the self energy. If we ignore \(\Lambda(\epsilon) \) and assume that \(W\gg \Gamma\), we can integrate Eq. \ref{eq:imppop} and \(\lambda_o \) can be approximated by
		\begin{align}\label{}
			\lambda_o \approx -\frac{\Gamma}{\pi} \ln(\frac{2W}{\Gamma})
		\end{align}
		Because we assume we are at the symmetric crossing point, 
		\begin{align}\label{}
			\lambda_v = -\lambda_o
		\end{align}
		
		Now, an eigenvalue of  \(\tilde{H}_{CIS}\), denoted \(\lambda\), must satisfy the self-consistent equation Eq. \ref{eq:chareq}:
		\begin{align}\label{eq:cischareq}
			\lambda = \sum_{\ta} \frac{\abs{\mel{\ta}{H}{d_v}}^2}{\lambda - (\lambda_{\ta} - \lambda_v)} + \sum_{\ti} \frac{\abs{\mel{\ti}{H}{d_o}}^2}{\lambda - (\lambda_o - \lambda_{\ti})}
		\end{align}
		As above, because we are at the symmetric crossing point, 
		the two terms on the right hand side of Eq. \ref{eq:cischareq} equal. Finally, our task is to find the smallest \(\lambda \) which satisfies the equation
		\begin{align}\label{eq:gap}
			\lambda = 2 \sum_{\ta} \frac{\abs{\mel{\ta}{H}{d_v}}^2}{\lambda - (\lambda_{\ta} - \lambda_v)}
		\end{align}
		
		To solve Eq. \ref{eq:gap}, we begin by noticing that, at the diabatic crossing, 
		\begin{align}\label{}
			\ket{d} = \frac{1}{\sqrt{2}} \left(\ket{d_o} + \ket{d_v} \right)
		\end{align}
		Moreover, for any Hamiltonian of the form in Eq. \ref{eq:H}, \(\ket{d}\) can be expanded in the full set of eigenstates\cite{Mahan2013}
		\begin{align}\label{}
			\ket{d} = \sum_{p} \frac{V}{\sqrt{\lambda_p^2 + (\Gamma/2)^2} } \ket{p}
		\end{align}
		If we further assume the rotated bath orbitals can be approximated by the original bath orbitals, say \(\ta \approx a \), then
		\begin{align}\label{eq:sclambda}
			\lambda &\approx 4 \sum_{a} \frac{V^2 \lambda_a^2}{\lambda_a^2 + (\Gamma/2)^2} \frac{1}{\lambda + \lambda_v - \lambda_a} = \frac{2\Gamma}{\pi} \int_0^W  \frac{\epsilon^2}{\epsilon^2 + (\Gamma/2)^2} \frac{1}{\lambda+\lambda_v-\epsilon} \dd \epsilon\nonumber \\
			&= -\frac{2\Gamma}{\pi} \ln\abs{\frac{r}{v}} -\frac{2\Gamma}{\pi} \frac{1}{v^2+1} \left(v\arctan(r) + \ln\abs{v} \right)
		\end{align}
		Here, in Eq. \ref{eq:sclambda} we have defined
		\begin{align}\label{}
			r &\equiv \frac{2W}{\Gamma}\\
			v & \equiv 2(\lambda+\lambda_v)/\Gamma
		\end{align}
		Recall \(\lambda_v \approx (\Gamma/\pi)\ln(2W/\Gamma)=(\Gamma/\pi)\ln(r) \), and denote
		\begin{align}\label{eq:defx}
			x \equiv \frac{\lambda+2\lambda_v}{\Gamma} = \frac{\lambda}{\Gamma} + \frac{2}{\pi}\ln(r)
		\end{align}
		so that
		\begin{align}\label{}
			v &= 2\left(x-\frac{1}{\pi}\ln(r) \right)
		\end{align}
		In the end, Eq. \ref{eq:sclambda} becomes (assuming \(r\gg 1\) so that \(\arctan(r) \approx \pi/2 \))
		\begin{align}\label{eq:x}
			x= \frac{2}{\pi} \frac{v^2}{v^2+1}\ln\abs{v}  - \frac{v}{v^2+1}
		\end{align}
		Eq. \ref{eq:x} is a transcendental equation for \(x\) with parameter \(r=2W/\Gamma\).
		While there is no easy analytical solution, it is obvious that the solution is on the order of 1 for any reasonable \(r\). In other words, from Eq. \ref{eq:defx}, \(\lambda + 2\lambda_v = \lambda + \lambda_v - \lambda_o \approx \Gamma \). 
	}
	\bibliography{Library}

\begin{mcitethebibliography}{106}
\bibitem[foo()]{footnote:AFSSH}
Of course, as is well known, this argument is not exact and will get worse with
  time because of different forces on different surfaces, eventually requiring
  a decoherence correction. For now, however, this nuance can be safely
  ignored.\relax
\mciteBstWouldAddEndPunctfalse
\mciteSetBstMidEndSepPunct{\mcitedefaultmidpunct}
{}{\mcitedefaultseppunct}\relax
\EndOfBibitem
\end{mcitethebibliography}
	
\end{document}